\documentclass[12pt]{article}
\usepackage{amsfonts, graphicx, verbatim, mathtools,amssymb, amsthm, mathrsfs}
\usepackage{tikz}
\usetikzlibrary{arrows.meta,positioning}
\usepackage{epstopdf}
\usepackage{graphicx}

\usepackage{epsfig,amsmath,amssymb,mathrsfs}
\usepackage[utf8]{inputenc}
\usepackage[colorlinks=true,citecolor=blue,,linktocpage=true,linkcolor=blue,urlcolor=black]{hyperref}
\usepackage{mathtools,slashed}
\usepackage{array}
\usepackage{xcolor}
\usepackage[force]{feynmp-auto}
\usepackage{caption}
\usepackage{subcaption}
\usepackage{overpic,youngtab}
\usepackage[latin,english]{babel}
\usepackage{amsmath}
\usepackage{amssymb}
\usepackage{epstopdf}
\usepackage{graphics,psfrag,rotating}
\usepackage{mathabx}
\usepackage{graphicx}
\usepackage{dcolumn}
\usepackage{float}
\usepackage{pdflscape}
\usepackage{array}
\usepackage{booktabs}
\usepackage{amscd}
\usepackage{mathtools}
\usepackage{fancybox}
\usepackage{fix-cm}
\usepackage[colorlinks=true,citecolor=blue,,linktocpage=true,linkcolor=blue,urlcolor=black]{hyperref}
\usepackage{tikz}
\usetikzlibrary{matrix}
\usetikzlibrary{positioning}
\usetikzlibrary{arrows}

  {\end{list}}%

\makeatletter
\renewcommand{\section}{\setcounter{equation}{0}\@startsection
 {section}%
 {1}%
 {0pt}%
 {-1\baselineskip}%
 {0.4\baselineskip}%
 {\bfseries\large}}%
\renewcommand{\subsection}{\@startsection
 {subsection}%
 {2}%
 {0pt}%
 {-0.75\baselineskip}%
 {0.2\baselineskip}%
 {\bfseries}}%
\renewcommand{\subsubsection}{\@startsection
 {subsubsection}%
 {3}%
 {0pt}%
 {-0.5\baselineskip}%
 {0.1\baselineskip}%
 {\sc}}%
\makeatother

\DeclareMathAlphabet{\mathpzc}{OT1}{pzc}{m}{it}

\usepackage{tikz}
\usetikzlibrary{patterns,snakes}
\usetikzlibrary{shapes.misc}
\usetikzlibrary{decorations.markings,decorations.pathmorphing}
\usetikzlibrary{calc}
\tikzstyle{spring}=[line width=0.8,black,snake=coil,segment amplitude=4.25,segment length=4.75,line cap=round]

\def\A{{\rm A}}

\def\be{\begin{equation}}
\def\ee{\end{equation}}

\def\g5{\gamma_{5}}

\def\id3k{\int\!\! \dfrac{d^3\!\vec{k}}{(2\pi)^3 }}

\def\tom{\tilde{\omega}}

\def\idpq{\int\!\! \dfrac{d^4\!p}{(2\pi)^4}\dfrac{d^4\!q}{(2\pi)^4}}

\def\idx{\int\!\! d^4\!x}

\def\idx{\int\!\! d^{4}\!x}

\newcommand{\bea}{\begin{eqnarray}}
\newcommand{\eea}{\end{eqnarray}}
\newcommand{\beann}{\begin{eqnarray*}}
\newcommand{\eeann}{\end{eqnarray*}}
\newcommand{\ba}{\begin{array}}
\newcommand{\ea}{\end{array}}

 \def\tp {\tilde{p}}

 \def\g {\gamma}

\newcommand{\email}[1]{\href{mailto:#1}{\tt #1}}

\begin{document}

\rightline{\scriptsize{FT/UCM 57-2025}}
\vglue 50pt

\begin{center}

{\LARGE \bf Entanglement through high-energy scattering in noncommutative quantum electrodynamics.}\\
\vskip 1.0true cm
{\Large Carmelo P. Martin}
\\
\vskip .7cm
{
	
	{Universidad Complutense de Madrid (UCM), Departamento de Física Teórica and IPARCOS, Facultad de Ciencias Físicas, 28040 Madrid, Spain}
	
	\vskip .5cm
	\begin{minipage}[l]{.9\textwidth}
		\begin{center}
			\textit{E-mail:}
			\email{carmelop@fis.ucm.es}.
			
		\end{center}
	\end{minipage}
}
\end{center}
\thispagestyle{empty}

\begin{abstract}
We analyze the tree-level generation of entanglement through some key scattering processes in massless  quantum electrodynamics on canonical noncomutative spacetime with space-space type of noncommutativity. The fermions in the noncommutative theory will be zero charge fermions.
The scattering processes we shall study do not occur in ordinary  Minkowski spacetime. We shall use the concurrence to characterize the amount of entanglement generated through a given scattering process. We shall show that, at tree-level, the concurrence for the scattering of two photons of opposite helicity is given by the same expression as in the case of the scattering of gluons in ordinary Minkowski spacetime.
Thus, maximal entanglement is achieved if and only if the polar scattering angle is equal to $\pi/2$. We also compute the concurrence for the head-on collision in the laboratory reference frame of two fermions of opposite helicity to obtain the same result as in the case of photon scattering.  Finally, we shall study a type of collision at right angles in the laboratory frame of fermions with opposite helicity. We show that in the latter case the concurrence depends on energy of the incoming fermions, the noncommutativity matrix $\theta^{ij}$, the polar, $\theta$, and azimuth angle, $\phi$, of the zero-momentum frame of the incoming fermions. In this latter case we see that when $\theta=\pi/2$ there are values of $\phi$ for which no entanglement is generated.
\end{abstract}

\vfill
\clearpage

\section{Introduction}

The Moyal (also called canonical) noncommutative spacetime is  obtained by assuming that the spacetime coordinates no longer commute but are self-adjoint operators --say $X^\mu, \mu=0,1,2,3$ being a Lorentz index-- satisfying the commutation relations
\begin{equation*}
[X^\mu,X^\nu]=i\omega^{\mu\nu}.
\end{equation*}
$\omega^{\mu\nu}$ is a constant antisymmetric matrix, called the noncommutativity matrix. The limit $\omega^{\mu\nu}\rightarrow 0$ yields, at least formally, ordinary Minkowski spacetime. The commutation relations just mentioned give rise to  Heisenberg-like uncertainty relations,
$\Delta X^\mu \Delta X^\nu \geq |\omega^{\mu\nu}|$,
 which set a minimum nonzero length, $l_{min}$, -the smallest $\sqrt{|\omega^{\mu\nu}|}$-- which can be effectively measured. Hence, the points of ordinary spacetime are to replaced by cells of non zero size  given by $l_{min}$ so that Moyal noncommutative spacetime is no longer made out of points. Having a nonzero minimum length that can be effectively probed is in harmony with the combination of quantum mechanics and the generation of black holes, for this combination leads to the conclusion that there must be a minimum nonzero value in the accuracy of the localization process of an event in spacetime \cite{Doplicher:1994tu}: probing regions of ever decreasing size demands an ever increasing energy, which eventually gives rise to a black hole and, hence, no information  would escape, classically, from the latter. So, it would appear that the classical notion of spacetime as a differentiable manifold would cease to make sense at high enough energies.

Canonical noncommutative spacetime arises naturally in the string theory framework.  It is  well known \cite{Ardalan:1998ce, Chu:1998qz, Schomerus:1999ug, Seiberg:1999vs} that when an open string ends on a D-brane in the presence of a constant Neveu-Schwartz $B$-field, the D-brane world volume becomes, upon quantization, a canonical noncommutative spacetime.

There is an equivalent definition of the Moyal noncommutative spacetime which is made by using the algebra, $\cal{A}_{\omega}$, of Schwartz functions on $\mathbb{R}^4$ with the $*$-product (also called Moyal)  of functions:
\begin{equation}
(f\star g)(x)= \idpq\,e^{-\frac{i}{2} p_\mu\omega^{\mu\nu}q_\nu}\;e^{-i(p+q)_\mu x^\mu}\;\tilde{f}(p)\tilde{g}(q).
\label{Moyalprod}
\end{equation}
The equivalence is established by using the Weyl transform -see \cite{Szabo:2001kg}, for details. For Euclidean signature and  by using $\cal{A}_{\omega}$, it has been shown in reference \cite{Gayral:2003dm}, that the canonical noncommutative space is an instance of noncommutative spin manifold in the sense of Connes's noncommutative geometry \cite{Connes:2019vcx} --see also \cite{Gracia-Bondia:2001upu}.

Gauge theories can be defined on canonical noncommutative spacetime -see \cite{Szabo:2001kg} and \cite{Hersent:2022gry}, for reviews. The gauge transformations being a deformation as given by $*$-product of functions of the ordinary gauge transformations. Actually, it was shown in reference \cite{Seiberg:1999vs} that there is limit in which
the low energy dynamics of the open string theory ending on a brane in the presence of a constant $B$-field is given by a Yang-Mills theory on canonical noncommutative spacetime with gauge symmetry being given by noncommutative $U(1)$ gauge transformations. This gauge theory on canonical noncommutative spacetime, which is called noncommutative $U(1)$ Yang-Mills theory,  is the counterpart of ordinary electrodynamics in the absence of matter. And yet, unlike in the ordinary case, noncommutative $U(1)$ gauge fields  --which give rise to photons-- do interact.

Dirac fermionic matter can be added to the noncommutative $U(1)$ Yang-Mills theory to obtain the noncommutative counterpart of ordinary quantum electrodynamics (QED). The noncommutative field theory in question is called noncommutative QED and, at tree-level, it has two salient features which are very relevant to the discussion carried out in this paper. First, photons do interact and, secondly,  fermion fields with zero electric charge couple to the photon field. These two features are consequences of the spacetime being a noncommutative manifold. Noncommmutative QED has been analyzed from several points of view: a partial list of references is constituted by items \cite{Hayakawa:1999yt} to \cite{Trampetic:2023qfv} in the section References.

Today, there is a lot of activity in the study of entanglement phenomena in high-energy physics --see references \cite{Blasone:2007vw} to \cite{Nunez:2025dch}, and references therein. In references
\cite{Cervera-Lierta:2017tdt} and \cite{Nunez:2025dch}, their authors study the tree-level generation of entanglement through the scattering of particles with well-defined helicity in QED and QCD; the degree of entanglement generated in a given scattering process being characterized by the concurrence.

Up to best of our knowledge, no study of the generation of entanglement in scattering processed in noncommutative gauge theories has been  displayed so far in the scientific literature. The purpose of this paper is to (partially) remedy  this unwanted state of affairs. We shall  follow in the footsteps of the authors of \cite{Cervera-Lierta:2017tdt} and \cite{Nunez:2025dch} and analyze the generation of entanglement in the tree-level scattering of photons and zero charge massless fermions in noncommutative QED (ie, QED on canonical noncommutative spacetime) for space-space type --i.e., $\omega^{0i}=0$ $\forall i=1,2,3$-- of noncommutativity. We shall consider tree-level scattering processes which do not happen in ordinary Minkowski spacetime, namely: the scattering of two photons with well-defined helicity and the scattering of fermions of opposite helicity. The amount of entanglement generated through these scattering processes will be characterized by the concurrence. We shall see that when the photons which collide have opposite helicity the concurrence has the same expression as in the case of gluons studied in \cite{Nunez:2025dch}. Hence maximal entanglement is generated if and only polar scattering angle, $\theta$, is equal to $\pi/2$. The same result is obtained if  fermions of opposite helicity collide head-on in the laboratory reference frame. And yet, we shall set up a nice type of scattering process where the incoming fermions collide at right angles in the laboratory reference frame and which yields a concurrence which depends on the energy of the incoming fermions, the noncommutativity matrix $\omega^{\mu\nu}$ and the polar, $\theta$, and azimuth, $\phi$, scattering angles --defined in the zero-momentum frame of the incoming photons. Actually, in this scattering process, when $\theta=\pi/2$, there are always values of $\phi$ for which the concurrence vanishes: no entanglement is generated.

Now, a point of terminology. Throughout this paper we shall call laboratory reference frame an inertial\footnote{By inertial we mean that the four momentum of a free particle does not change.} reference system in which momenta of the particles are measured and --what is key-- where  the noncommutativity matrix,  $\omega^{\mu\nu}$, in (\ref{Moyalprod}) is such that $\omega^{0i}=0,\quad\forall i=1,2,3$. Of course, by applying a Lorentz transformation to the  laboratory reference frame  we shall get another reference frame where  the momentum coordinates and $\omega^{\mu\nu}$ are Lorentz transformed without changing the physics coming from the noncommutative field theory. This a passive Lorentz transformation. However,  it is clear that an active -- the Lorentz transformation changes the momenta but not $\omega^{\mu\nu}$-- is not a symmetry of the noncommutative field theory. In this paper we shall also consider the zero-momentum frame of two colliding particles, which is obtained from to the laboratory reference frame an appropriate Lorentz transformation.

The lay out of this paper is as follows. In section 2 we put forward cursorily the formalism needed to carry out in the computations done in the next sections. In section 3 we analize the generation of entanglement through photon-photon scattering. Section 4 is devoted to the study of the tree-level scattering of zero charge massless fermions. Here we have used Mathematica \cite{mathe} to do some numerical computations.
Our conclusions can be found in section 5.

\section{Noncommutative QED with massless zero charge fermions}

Let $A_\mu(x)$ stand for a $U(1)$ gauge field on canonical noncommutative spacetime. Under noncommutative  infinitesimal $U(1)$ gauge transformation, $\chi(x)$, the field $A_\mu(x)$ changes as follows: $\delta A_\mu(x)=ie[A_\mu,\chi]_{\star}(x)=ie(A_\mu\star\chi(x)-\chi\star\A_\mu(x))$. Let $\psi(x)$ denote a Dirac fermion on canonical noncommutative spacetime, which transforms under the previous infinitesimal $U(1)$ transformations as follows: $\delta \psi(x)=ie[\psi,\chi]_{\star}(x)=ie(\psi\star\chi(x)-\chi\star\psi(x))$. Them, the  action of noncommutative QED with a massless zero charge fermion reads\footnote{Let us mention that the massive version of this model has been used in \cite{Schupp:2002up, Minkowski:2003jg, Horvat:2011iv, Horvat:2017aqf} to  describe the propagation and interaction of neutrinos on canonical noncommutative space-time.}
\begin{equation}
S=-\cfrac{1}{4}\,\idx\;F_{\mu\nu}(x) F^{\mu\nu}(x)+i\idx\;\bar{\psi}(x)\gamma^\mu D_{\mu}[A]\psi(x),
\label{action}
\end{equation}
where $F_{\mu\nu}(x)=\partial_\mu A_\nu(x)-\partial_\nu A_\mu(x)-ie [A_\mu,A_\nu]_{\star}(x)$ and $D_{\mu}[A]\psi(x)=\partial_\mu\psi(x)-ie[A_\mu,\psi]_{\star}(x)$.
The symbol $[\phantom{f} ,\phantom{g}]_{\star}$ shows that the commutator is defined with regard to the $\star$-product in (\ref{Moyalprod}).

In this paper we shall assume that $\omega^{\mu\nu}$ in  (\ref{Moyalprod}) is such that $\omega^{0i}=0$ $\forall i=1,2,3$. The  reason for this choice is twofold: $\it{i})$ the standard formalism as defined in terms of Feynman diagrams is not consistent \cite{Gomis:2000zz} with unitarity if $\omega^{0i}\neq 0$ and $\it{ii})$ that field theories for $\omega^{0i}=0$ $\forall i=1,2,3$ are the ones which arise naturally in string theory --see \cite{Alvarez-Gaume:2001dfr}, for further discussion.

In the sequel, we shall need the following Feynman rules derived from the action in (\ref{action}):
\begin{equation}
\begin{array}{l}
{\langle A_\mu(p) A_\nu(-p)\rangle=\cfrac{-i\eta_{\mu\nu}}{p^2},}\\[8pt]
{V^{\text{AAA}}_{\mu_1\mu_2\mu_3}(p_1,p_2,p_3)=
2e\sin\big(\cfrac{1}{2}\,p_1\omega p_2\big)\,V^{(3)}_{\mu_1\mu_2\mu_3}(p_1,p_2,p_3)}\\[8pt]
{V^{(3)}_{\mu_1\mu_2\mu_3}(p_1,p_2,p_3)=
(p_1-p_2)_{\mu_3}\eta_{\mu_1\mu_2}+(p_2-p_3)_{\mu_1}\eta_{\mu_2\mu_3}+
(p_3-p_1)_{\mu_2}\eta_{\mu_3\mu_1},}\\[8pt]
{V^{\text{AAAA}}_{\mu_1\mu_2\mu_3\mu_4}(p_1,p_2,p_3,p_4)=4ie^2\big[\sin\big(\cfrac{1}{2}\,p_1\omega p_2\big)\sin\big(\cfrac{1}{2}\,p_3\omega p_4\big)(\eta_{\mu_1\mu_4}\eta_{\mu_2\mu_3}-
\eta_{\mu_1\mu_3}\eta_{\mu_2\mu_4})+}\\[4pt]
{\phantom{V^{\text{AAAA}}_{\mu_1\mu_2\mu_3\mu_4}(p_1,p_2,p_3,p_4)=4ie^2\big[}
\sin\big(\cfrac{1}{2}\,p_3\omega p_1\big)\sin\big(\cfrac{1}{2}\,p_2\omega p_4\big)(\eta_{\mu_1\mu_3}\eta_{\mu_2\mu_4}-
\eta_{\mu_1\mu_2}\eta_{\mu_3\mu_4})+}\\[4pt]
{\phantom{V^{\text{AAAA}}_{\mu_1\mu_2\mu_3\mu_4}(p_1,p_2,p_3,p_4)=4ie^2\big[}
\sin\big(\cfrac{1}{2}\,p_1\omega p_4\big)\sin\big(\cfrac{1}{2}\,p_2\omega p_3\big)(\eta_{\mu_1\mu_2}\eta_{\mu_3\mu_4}-
\eta_{\mu_1\mu_4}\eta_{\mu_2\mu_3})\big],}\\[8pt]
{V^{\bar{\psi}\rm{A}\psi}_{\mu}(p_1,p_3,p_2)=2ie\sin\big(\cfrac{1}{2}\,p_1\omega p_2\big)\gamma_\mu.}
\label{Feynrules}
\end{array}
\end{equation}
We use the notation $p\omega q\equiv p_\mu\omega^{\mu\nu}q_\nu$. Our Minkowski metric is a mostly plus metric.

\section{Entanglement through $\gamma\gamma\rightarrow \gamma\gamma$ scattering}

The scattering of two photons with momentum and helicities being $(p_1,h_1\, 1)$ and $(p_2,h_2\, 1)$, $h_1=\pm$ and $h_2=\pm $, respectively, in the laboratory reference frame gives rise to the transition
\begin{equation}
|i\rangle\rightarrow|f\rangle
\label{scatpro}
\end{equation}
where
\begin{equation*}
\begin{array}{l}
{|i\rangle = |(p_1,h_1)(p_2,h_2)\rangle,}\\[8pt]
{|f\rangle = C[h_1h_2;++]\,|(p_3,+)(p_4,+)\rangle+C[h_1h_2;+-]\,|(p_3,+)(p_4,-)\rangle+}\\[4pt]
{\phantom{|f\rangle = }
C[h_1h_2;-+]\,|(p_3,-)(p_4,+)\rangle+C[h_1 h_2;--]|(p_3,-)(p_4,-)\rangle.}
\end{array}
\end{equation*}
Note that $p_1$ and $p_2$ are incoming momenta, whereas $p_3$ and $p_4$ are outgoing momenta. The coefficients $C[h_1h_2;++]$, $C[h_1h_2;+-]$, $C[h_1h_2;-+]$ and $C[h_1h_2;--]$ in the previous expression are obtained by computing the following ${\cal M}$-matrix elements
\begin{equation}
\begin{array}{l}
{{\cal M}[h_1 h_2; ++]={\cal M}[(p_1,h_1),(p_2,h_2);(p_3,+),(p_4,+)],}\\[4pt]
{{\cal M}[h_1 h_2; +-]={\cal M}[(p_1,h_1),(p_2,h_2);(p_3,+),(p_4,-)]}\\[4pt]
{{\cal M}[h_1 h_2; -+]={\cal M}[(p_1,h_1),(p_2,h_2);(p_3,-),(p_4,+)],}\\[4pt]
{{\cal M}[h_1 h_2; --]={\cal M}[(p_1,h_1),(p_2,h_2);(p_3,-),(p_4,-)].
}
\label{Melements}
\end{array}
\end{equation}
Indeed, we have
\begin{equation*}
\begin{array}{l}
{C[h_1h_2;++]=\cfrac{1}{\cal N}\;{\cal M}[h_1 h_2; ++],\quad C[h_1h_2;+-]=\cfrac{1}{\cal N}\;{\cal M}[h_1 h_2; +-],}\\[4pt]
{C[h_1h_2;-+]=\cfrac{1}{\cal N}\;{\cal M}[h_1 h_2; -+],\quad C[h_1h_2;--]=\cfrac{1}{\cal N}\;{\cal M}[h_1 h_2; --],}\\[6pt]
{{\cal N}=\Big(|{\cal M}[h_1 h_2; ++]|^2+|{\cal M}[h_1 h_2; +-]|^2+|{\cal M}[h_1 h_2; -+]|^2+|{\cal M}[h_1 h_2; --]|^2\Big)^{1/2}.
}
\end{array}
\end{equation*}

To characterize the amount of entanglement generated by the scattering process in (\ref{scatpro}), we shall use, as in \cite{Cervera-Lierta:2017tdt, Nunez:2025dch}, the concurrence, which is denoted by $\Delta$. $\Delta$ is defined as follows
\begin{equation}
\Delta=2|C[h_1h_2;++]\; C[h_1h_2;--]-C[h_1h_2;+-]\;C[h_1h_2;-+]|
\label{concurrence}
\end{equation}
It can be shown that the concurrence satisfies $0\leq \Delta\leq 1$. $\Delta=0$ signals no entanglement, $\Delta=1$ implies maximal entanglement.

By explicit computation, we have shown that if both $h_1$ and $h_2$ are $+$, then, $C[++;+-]$, $C[++;-+]$ and $C[++;--]$ vanish at tree-level, so that no tree-level entanglement occurs. Analogously, if
both $h_1$ and $h_2$ are $-$, then, no tree-level entanglement  arise either, for $C[--;++]$, $C[--;+-]$ and $C[--;-+]$ are zero at tree-level. We will not give the details of the standard type of computations that lead to the previous conclusions, but, instead, we shall use known results -see reference \cite{Badger:2023eqz}-- obtained by using the spinor helicity formalism to argue that only if the helicities of the incoming photons are such $(h_1,h_2)=(+,-)$ or $(h_1,h_2)=(-,+)$, the scattering process in (\ref{scatpro}) yields entanglement. Indeed, let us first recall that, due to crossing symmetry \cite{Srednicki:2007qs}, any amplitude involving two incoming and two outgoing particles is related to the amplitude of four incoming particles upon replacing the former two outgoing particles with their antiparticles with opposite momentum and opposite helicity. Of course, this relation is such that if the latter amplitude vanish, so does the corresponding amplitude with two incoming and two outgoing particles. Now, it has been shown in reference \cite{Huang:2010fc} that any tree-level amplitude involving four incoming photons on canonical noncommutative spacetime can be expressed as follows
\begin{equation*}
\begin{array}{l}
{{\cal M}[(k_1,h_1),(k_2,h_2),(k_3,h_3),(k_4,h_4)]=}\\[4pt]
{\sum\limits_{\sigma\in S_3}\, \phi(k_1,k_{\sigma(2)},k_{\sigma(3}),k_{\sigma(4)})\;
\tilde{\cal M}[(k_1,h_1),(k_{\sigma(2)},h_{\sigma(2)}),(k_{\sigma(3)},h_{\sigma(3)}),(k_{\sigma(4)},h_{\sigma(4))}],}
\end{array}
\end{equation*}
where
\begin{equation*}
\phi(q_1,q_2,q_3,q_4)= e^{-\frac{i}{2}\big(\sum\limits_{1\leq i< j\leq 4}\ q^\mu_i\omega_{\mu\nu}q^\nu_j\big)}
\end{equation*}
and $\tilde{\cal M}[(q_1,h_1),(q_2,h_2),(q_3,h_3,(q_4,h_4)]$ is the standard colour-ordered tree-level amplitude for four incoming  gluons. It is well known \cite{Badger:2023eqz} that
$\tilde{\cal M}[(q_1,h_1),(q_2,h_2),(q_3,h_3,(q_4,h_4)]$ vanishes when the four helicities have the same value or when three of them have also the same value. On the other hand, if two of  the helicities are equal and have opposite value to the other two, then, $\tilde{\cal M}[(q_1,h_1),(q_2,h_2),(q_3,h_3,(q_4,h_4)]$ does not vanish and it is given by the famous Parke-Taylor formula. Hence, we conclude that
\begin{equation*}
\begin{array}{l}
{{\cal M}[(k_1,+),(k_2,+),(k_3,+),(k_4,+)]=0, {\cal M}[(k_1,-),(k_2,-),(k_3,-),(k_4,-)]=0,}\\[4pt]
{{\cal M}[(k_1,-),(k_2,+),(k_3,+),(k_4,+)]=0,{\cal M}[(k_1,+),(k_2,-),(k_3,+),(k_4,+)]=0,}\\[4pt]
{{\cal M}[(k_1,+),(k_2,+),(k_3,-),(k_4,+)]=0, {\cal M}[(k_1,+),(k_2,+),(k_3,+),(k_4,-)]=0,}\\[4pt]
{{\cal M}[(k_1,+),(k_2,-),(k_3,-),(k_4,-)]=0,{\cal M}[(k_1,-),(k_2,+),(k_3,-),(k_4,-)]=0,}\\[4pt]
{{\cal M}[(k_1,-),(k_2,-),(k_3,+),(k_4,-)]=0,{\cal M}[(k_1,-),(k_2,-),(k_3,-),(k_4,+)]=0.}\\[4pt]
\end{array}
\end{equation*}
As we pointed out above, these results and  crossing symmetry imply that the only non-zero tree-level matrix elements in (\ref{Melements}) are
\begin{equation*}
\begin{array}{l}
{{\cal M}[+ +; ++]={\cal M}[(p_1,+),(p_2,+);(p_3,+),(p_4,+)],}\\[4pt]
{{\cal M}[+-; +-]={\cal M}[(p_1,+),(p_2,+);(p_3,+),(p_4,-)],}\\[4pt]
{{\cal M}[+-; -+]={\cal M}[(p_1,+),(p_2,-);(p_3,-),(p_4,+)],}\\[4pt]
{{\cal M}[--; --]={\cal M}[(p_1,-),(p_2,-);(p_3,-),(p_4,-)],}\\[4pt]
{{\cal M}[-+; +-]={\cal M}[(p_1,-),(p_2,+);(p_3,+),(p_4,-)]}\\[4pt]
{{\cal M}[-+; -+]={\cal M}[(p_1,-),(p_2,+);(p_3,-),(p_4,+)].
}
\end{array}
\end{equation*}
Recall that $p_1$ and $p_2$ are incoming momenta and $p_3$ and $p_4$ are outgoing momenta. As we said above, we have verified by explicit computation that indeed those are the only non-vanishing matrix elements. Hence, we conclude that entanglement only arises in the following processes
\begin{equation}
\begin{array}{l}
{|(p_1,+)(p_2,-)\rangle\rightarrow C[+-;+-]\,|(p_3,+)(p_4,-)\rangle+C[+-;-+]\,|(p_3,-)(p_4,+)\rangle,}\\[4pt]
{|(p_1,-)(p_2,+)\rangle\rightarrow C[-+;+-]\,|(p_3,+)(p_4,-)\rangle+C[-+;-+]\,|(p_3,-)(p_4,+)\rangle.}
\label{entprocess}
\end{array}
\end{equation}
Obviously, nothing new will be learned from the second process, once we have analyzed the generation of entanglement through first process.

According to (\ref{concurrence}), the concurrence for the first process in (\ref{entprocess}) reads
\begin{equation}
\Delta=2\;\frac{|{\cal M}[+-;+-]|\,|{\cal M}[+-;-+]|}{|{\cal M}[+-;+-]|^2+|{\cal M}[+-;-+]|^2}.
\label{concurrencephotons}
\end{equation}

Let us focus now on the computation of ${\cal M}[+-;-+]$ and ${\cal M}[+-;-+]$ at tree-level. These amplitudes can be computed by summing the contribution of the tree-level Feynman diagrams corresponding to the $s,t$ and $u$ channels plus the contribution coming from the four-photon vertex, $V^{\text{AAAA}}_{\mu_1\mu_2\mu_3\mu_4}(p_1,p_2,p_3,p_4)$, in (\ref{Feynrules}). Thus, we have
\begin{equation}
\begin{array}{l}
{{\cal M}[+-;h_3 h_4]={\cal M}_s[+-; h_2;h_3 h_4]+{\cal M}_t[+-;h_3 h_4]+{\cal M}_u[+-;h_3 h_4]+{\cal M}_4[+-;h_3 h_4],}\\[8pt]
{{\cal M}_s[+-;h_3 h_4]=-4e^2\,\sin[\frac{1}{2}p_1\omega p_2]\sin[\frac{1}{2}p_3\omega p_4]\;\varepsilon^{\mu_1}(p_1,+)\varepsilon^{\mu_2}(p_2,-)\times}\\
{\phantom{{\cal M}_s[+-;}
V^{(3)}_{\mu_1\mu_2\mu}(p_1,p_2,-p_1-p_2)\cfrac{\eta^{\mu\nu}}{s}V^{(3)}_{\nu\mu_3\mu_4}(p_3+p_4,-p_3,-p_4)
(\varepsilon^{\mu_3}(p_3,h_3))^{*}(\varepsilon^{\mu_4}(p_4,h_4))^{*},}\\[8pt]
{{\cal M}_t[+-;h_3 h_4]=-4e^2\,\sin[\frac{1}{2}p_1\omega p_3]\sin[\frac{1}{2}p_2\omega p_4]\;\varepsilon^{\mu_1}(p_1,+)\varepsilon^{\mu_2}(p_2,-)\times}\\
{\phantom{{\cal M}_s[+-;}
V^{(3)}_{\mu_1\mu_3\mu}(p_1,p_3,-p_1+p_3)\cfrac{\eta^{\mu\nu}}{t}V^{(3)}_{\nu\mu_2\mu_4}(-p_2+p_4,p_2,-p_4)
(\varepsilon^{\mu_3}(p_3,h_3))^{*}(\varepsilon^{\mu_4}(p_4,h_4))^{*},}\\[8pt]
{{\cal M}_u[+-;h_3 h_4]=-4e^2\,\sin[\frac{1}{2}p_1\omega p_4]\sin[\frac{1}{2}p_2\omega p_3]\;\varepsilon^{\mu_1}(p_1,+)\varepsilon^{\mu_2}(p_2,-)\times}\\
{\phantom{{\cal M}_s[+-;}
V^{(3)}_{\mu_1\mu_4\mu}(p_1,p_4,-p_1+p_4)\cfrac{\eta^{\mu\nu}}{u}V^{(3)}_{\nu\mu_2\mu_3}(-p_2+p_3,p_2,-p_3)
(\varepsilon^{\mu_3}(p_3,h_3))^{*}(\varepsilon^{\mu_4}(p_4,h_4))^{*},}\\[8pt]
{{\cal M}_4[+-;h_3 h_4]=\varepsilon^{\mu_1}(p_1,+)\varepsilon^{\mu_2}(p_2,-)V^{\text{AAAA}}_{\mu_1\mu_2\mu_3\mu_4}(p_1,p_2,p_3,p_4)
(\varepsilon^{\mu_3}(p_3,h_3))^{*}(\varepsilon^{\mu_4}(p_4,h_4))^{*}.
}
\label{amplitudesphotonphoton}
\end{array}
\end{equation}
where $h_3,h_4$ are equal to $\pm$, as needs be, and the vertex functions are to be found in (\ref{Feynrules}). Also recall that $p_1+p_2=p_3+p_4$ and that we use the notation
\begin{equation*}
p\omega q= p_\mu \omega^{\mu\nu} q_\nu.
\end{equation*}
The photon polarizations are defined as follows: given a photon with four-momenta $p^\mu=E(1,\cos\phi \sin\theta,\sin\phi\sin\theta,\cos\theta)$ and helicity $h$, its polarization is given by
\begin{equation*}
\varepsilon^\mu(p,h)=(0,\vec{\varepsilon}\,(p,h)),\;
\vec{\varepsilon}\,(p,h)=\frac{e^{-i h \phi}}{\sqrt{2}}(\cos\theta \cos\phi + i h \sin\phi,\cos\theta \sin\phi - i h \cos\phi, -\sin\theta).
\end{equation*}

By applying the spinor helicity formalism to the expression above, one can easily show that
\begin{equation*}
{\cal M}_s[+-;+-]=0={\cal M}_s[+-;-+].
\end{equation*}

To make easier the comparison with the final result in \cite{Nunez:2025dch}, we shall change the momenta and the noncommutativity matrix $\omega^{\mu\nu}$ to the zero-momentum reference frame of the incoming photons. This change is implemented by
\begin{equation}
\begin{array}{l}
{p^\mu_1\rightarrow \tilde{p}^\mu_1= L^\mu_{\phantom{\mu}\nu}\, p^\nu_1,\quad p^\mu_2\rightarrow \tilde{p}^\mu_2= L^\mu_{\phantom{\mu}\nu}\, p^\nu_2,\quad
\omega^{\mu\nu}\rightarrow\tilde{\omega}^{\mu\nu}=L^\mu_{\phantom{\mu}\rho}\,L^\mu_{\phantom{\mu}\sigma}\,\omega^{\rho\sigma},}\\[8pt]
{L^0_{\phantom{0}0}=\gamma_{v},\quad L^i_{\phantom{i}0}=-\gamma_{v}\,v^i,\quad L^0_{\phantom{0}i}=\gamma_{v}\,v_i,\quad
L^i_{\phantom{i}j}=\delta^i_j+\frac{1-\gamma_{v}}{\vec{v}^2}\,v^i\,v_j}\\[8pt]
{\gamma_{v}=\frac{1}{\sqrt{1-\vec{v}^2}},\quad\vec{v}=\frac{1}{E_1+E_2}\,(\vec{p}_1+\vec{p}_2),\quad v_i=-v^i,\quad i=1,2,3,}
\label{zeromomentum}
\end{array}
\end{equation}
where $\tilde{p}_1^\mu$,  $\tilde{p}_2^{\mu}$ and $\tilde{\omega}^{\mu\nu}$ denote the incoming momenta and the noncommutativity matrix in the aforementioned zero-momentum reference frame. We can always take
\begin{equation}
\tilde{p}^\mu_1=\tilde{E}(1,0,0,1),\quad\tilde{p}^\mu_1=\tilde{E}(1,0,0,-1).
\label{inmomenta}
\end{equation}
Then, in the zero-momentum reference frame, the outgoing photons will have the following momenta
\begin{equation}
\tilde{p}^\mu_3=\tilde{E}(1,\cos\phi \sin\theta,\sin\phi \sin\theta,\cos\theta),\quad\tilde{p}^\mu_4=\tilde{E}(1,-\cos\phi \sin\theta,-\sin\phi \sin\theta,-\cos\theta),
\label{outmomenta}
\end{equation}
where $\theta\in[0,\pi]$ and $\phi\in[0,2\pi)$ are the polar and azimuth angles with regard to the direction defined by the three-vector of $(\tilde{p}^1_1,\tilde{p}^2_1,\tilde{p}^3_1)$.

It is plain that
\begin{equation*}
p_a^\mu\omega_{\mu\nu} p_b^\nu=\tilde{p}_a^\mu\tilde{\omega}_{\mu\nu}\tilde{p}_b^\nu\quad \forall a,b=1,2,3,4.
\end{equation*}

After taking into account the definitions in (\ref{amplitudesphotonphoton}), one gets, in the case at hand, the following results
\begin{equation*}
\begin{array}{l}
{{\cal M}_t[+-;+-]=-4e^2\sin[\frac{1}{2}\tp_1\tom \tp_3]\sin[\frac{1}{2}(\tp_1\tom \tp_2+\tp_2\tom \tp_3)]\,e^{-2 i\phi}\cos^4[\frac{\theta}{2}]\,\cfrac{3+\cos\theta}{\cos\theta-1},}\\[8pt]
{{\cal M}_u[+-;+-]=-4e^2\sin[\frac{1}{2}\tp_2\tom \tp_3]\sin[\frac{1}{2}(\tp_1\tom \tp_3-\tp_1\tom \tp_2)]\,e^{-2 i\phi}\cfrac{1}{4}(\cos\theta-3)(1+\cos\theta),}\\[8pt]
{{\cal M}_4[+-;+-]=4e^2\big(\sin[\frac{1}{2}\tp_1\tom \tp_3]\sin[\frac{1}{2}(\tp_1\tom \tp_2+\tp_2\tom \tp_3)]+}\\
{\phantom{{\cal M}_4[+-;+-]=4e^2\big(}
\sin[\frac{1}{2}\tp_2\tom \tp_3]\sin[\frac{1}{2}(\tp_1\tom \tp_3-\tp_1\tom \tp_2)]\big)
\,e^{-2 i\phi}\cos^4[\frac{\theta}{2}],}\\[8pt]
{{\cal M}_t[+-;-+]=-4e^2\sin[\frac{1}{2}\tp_1\tom \tp_3]\sin[\frac{1}{2}(\tp_1\tom \tp_2+\tp_2\tom \tp_3)]\,e^{-2 i\phi}\cfrac{1}{4}\,(3+\cos\theta)(\cos\theta-1),}\\[8pt]
{{\cal M}_u[+-;-+]=-4e^2\sin[\frac{1}{2}\tp_2\tom \tp_3]\sin[\frac{1}{2}(\tp_1\tom \tp_3-\tp_1\tom \tp_2)]\,e^{-2 i\phi}\sin^4[\frac{\theta}{2}]\,\cfrac{-3+\cos\theta}{\cos\theta+1},}\\[8pt]
{{\cal M}_4[+-;-+]=4e^2\big(\sin[\frac{1}{2}\tp_1\tom \tp_3]\sin[\frac{1}{2}(\tp_1\tom \tp_2+\tp_2\tom \tp_3)]+}\\
{\phantom{{\cal M}_4[+-;+-]=4e^2\big(}
\sin[\frac{1}{2}\tp_2\tom \tp_3]\sin[\frac{1}{2}(\tp_1\tom \tp_3-\tp_1\tom \tp_2)]\big)
\,e^{-2 i\phi}\sin^4[\frac{\theta}{2}],
}
\end{array}
\end{equation*}
Using the previous results, one obtains the following amplitudes
\begin{equation}
\begin{array}{l}
{{\cal M}[+-;+-]=\big[{\cal C}_1+{\cal C}_2+({\cal C}_1-{\cal C}_2)\,\cos\theta\big]e^{-2 i\phi}\cot[\theta/2],}\\[8pt]
{{\cal M}[+-;+-]=\big[{\cal C}_1+{\cal C}_2+({\cal C}_1-{\cal C}_2)\,\cos\theta\big]e^{-2 i\phi}\tan[\theta/2],
}
\label{photonphotonamplitude}
\end{array}
\end{equation}
with
\begin{equation}
\begin{array}{l}
{{\cal C}_1=4e^2\sin[\frac{1}{2}\tp_1\tom \tp_3]\sin[\frac{1}{2}(\tp_1\tom \tp_2+\tp_2\tom \tp_3)]}\\[8pt]
{{\cal C}_2=4e^2\sin[\frac{1}{2}\tp_2\tom \tp_3]\sin[\frac{1}{2}(\tp_1\tom \tp_3-\tp_1\tom \tp_2)].
}
\label{caligraphicces}
\end{array}
\end{equation}

It is plane that ${\cal C}_1$ and ${\cal C}_2$ depend on  the noncommutativity matrix\footnote{Trough $\tom^{\mu\nu}$}, $\omega^{\mu\nu}$,  and the angle $\phi$ as befits the fact that theory is no invariant under active Lorentz transformations. And yet, the concurrence obtained from (\ref{photonphotonamplitude}) by using (\ref{concurrencephotons}) reads
\begin{equation*}
\Delta=\frac{2\tan^4[\theta/2]}{1+\tan^8[\theta/2]},
\end{equation*}
which is the same concurrence --see \cite{Nunez:2025dch}-- as that of the tree-level scattering of gluons in ordinary Minkowski spacetime. Hence, in noncommutative QED maximal entanglement is obtained if and only if $\theta=\pi/2$.

\section{Entanglement through fermion-fermion$\rightarrow$ fermion-fermion scattering}

Let us now compute the concurrence for the  scattering process, on noncommutative space-time, of two incoming massless fermions of zero $U(1)$-charge and  momentum and helicities $(p_1,+1/2)$ and $(p_2,-1/2)$, respectively, in the laboratory reference frame, yielding two massless zero-charge outgoing fermions of momenta $p_3$ and $p_4$ and helicities $\pm 1/2$. This is the process below:
\begin{equation}
\begin{array}{l}
{|i\rangle = |(p_1,+1/2)(p_2,-1/2)\rangle,}\\[8pt]
{|f\rangle = C[+-;++]\,|(p_3,+1/2)(p_4,+1/2)\rangle+C[+-;+-]\,|(p_3,+1/2)(p_4,-1/2)\rangle+}\\[4pt]
{\phantom{|f\rangle = }
C[+-;-+]\,|(p_3,-1/2)(p_4,+1/2)\rangle+C[+-;--]|(p_3,-1/2)(p_4,-1/2)\rangle,}
\label{ferprocess}
\end{array}
\end{equation}
Let us introduce the following amplitudes
\begin{equation*}
\begin{array}{l}
{{\cal M}[+-;h_3 h_4]={\cal M}_t[+-;h_3 h_4]+{\cal M}_u[+-;h_3 h_4],}\\[8pt]
{{\cal M}_t[+-;h_3 h_4]=4 e^2 \sin[\frac{1}{2}p_1\omega p_2]\sin[\frac{1}{2}p_2\omega p_4]\,\frac{1}{t}\bar{u}(p_3,h_3)\gamma^\mu u(p_1,+)\bar{u}(p_4,h_4)\gamma^\mu u(p_2,-),}\\[8pt]
{{\cal M}_u[+-;h_3 h_4]=-4 e^2 \sin[\frac{1}{2}p_1\omega p_4]\sin[\frac{1}{2}p_2\omega p_3]\,\frac{1}{u}\bar{u}(p_4,h_4)\gamma^\mu u(p_1,+)\bar{u}(p_3,h_3)\gamma^\mu u(p_2,-).}
\end{array}
\end{equation*}
Then, the coefficients in (\ref{ferprocess}) are given by
\begin{equation}
\begin{array}{l}
{C[+-;++]=\cfrac{1}{\cal N}\;{\cal M}[+-; ++],\quad C[+-;+-]=\cfrac{1}{\cal N}\;{\cal M}[+-; +-],}\\[4pt]
{C[+-;-+]=\cfrac{1}{\cal N}\;{\cal M}[+-; -+],\quad C[+-;--]=\cfrac{1}{\cal N}\;{\cal M}[+-; --],}\\[6pt]
{{\cal N}=\Big(|{\cal M}[+-; ++]|^2+|{\cal M}[+-; +-]|^2+|{\cal M}[+-; -+]|^2+|{\cal M}[+-; --]|^2\Big)^{1/2}.
}
\label{ces}
\end{array}
\end{equation}

Let us change to the zero-momentum reference frame as we did in the previous section -see (\ref{zeromomentum}), (\ref{inmomenta}) and (\ref{outmomenta}). Then, one obtains the following results:
\begin{equation*}
\begin{array}{l}
{{\cal M}_t[+-;+-]=2\,{\cal C}_1\,e^{ i\phi}\cot^2[\theta/2],}\\[8pt]
{{\cal M}_u[+-;+-]=0,}\\
{{\cal M}_t[+-;-+]=0,}\\[8pt]
{{\cal M}_u[+-;-+]=2\,{\cal C}_2\,e^{ i\phi}\tan^2[\theta/2],}\\[8pt]
{{\cal M}_t[+-;++]=0,\quad{\cal M}_u[+-;++]=0,\quad{\cal M}_t[+-;--]=0,\quad{\cal M}_u[+-;--]=0,
}
\end{array}
\end{equation*}
where ${\cal C}_1$ and ${\cal C}_2$ are given in (\ref{caligraphicces}).

Let us recall that the concurrence, $\Delta$, is given by
\begin{equation*}
\Delta=2|C[+-;++]\; C[+-;--]-C[+-;+-]\;C[+-;-+]|.
\end{equation*}
Then, taking into account the previous results, (\ref{ces}), one gets
\begin{equation}
\Delta=\frac{2|{\cal C}_1{\cal C}_2|}{({\cal C}_1 \cot^2[\theta/2])^2+({\cal C}_2 \tan^2[\theta/2])^2}
\label{ferconcurrence}
\end{equation}

Due to the presence of terms of the type $\sin[\frac{1}{2}p\omega q]$ -which break invariance under active Lorentz transformations--  in ${\cal C}_1$ and ${\cal C}_2$ in (\ref{caligraphicces}), to compute the concurrence, $\Delta$, for arbitrary values of the four-momenta will lead to not very illuminating expressions. This is why we have decided to carry out the computations for two particular sets of scattering setups, namely, head-on collision  and collision at right angles of fermions in the laboratory reference frame.  Let us begin with the head-on collision case.

\subsection{Head-on collision of fermions}

In this case the three-momenta $\vec{p}_1$ and $\vec{p}_2$, the momenta in the laboratory reference frame, satisfy $\vec{p}_1=-\vec{p}_2$, so that the laboratory reference system is the zero-momentun reference frame of the fermions, ie, the Lorentz transformation in
(\ref{zeromomentum}) is the identity transformation. Hence, $\tp_1^{\mu}\tom_{\mu\nu}\tp_2^\nu=0$\footnote{recall that $\omega^{0i}=0$}, so that ${\cal C}_1={\cal C}_2$; which leads to
\begin{equation*}
\Delta=\frac{2 \tan^4[\theta/2]}{1+\tan^8[\theta/2]}.
\end{equation*}
In summary, when the fermions collide head-on the concurrence does not depend on the value of the noncommutativity matrix and there is maximal entanglement if and only if $\theta=\pi/2$.
Of course, the scattering process we have analyzed does not exist in ordinary Minkowski space, since in this spacetime our fermions would be free.

\subsection{Fermion scattering at right angles.}

Here we shall set up a class of scattering process of two photons where in zero-momentum reference frame  the concurrence depends of energy,  the noncommutativity matrix, $\omega^{ij}$ and the polar and azimuth angles of one of the outgoing photons.

Let us begin by parameterizing the noncommutative matrix $\omega^{\mu\nu}$ in (\ref{action}), and hence in (\ref{Feynrules}),  as follows:
\begin{equation*}
\omega^{\mu\nu}=\cfrac{c^{\mu\nu}}{\Lambda^2_{\text{nc}}},
\end{equation*}
were $c^{\mu\nu}$ is a dimensionless antisymmetric real matrix such that $c^{0i}=0$, $\forall i=1,2,3$, and $(c^{12})^2+(c^{13})^2+(c^{23})^2=3$. $\Lambda_{\text{nc}}$ sets the scale of noncommutativity and has dimensions of energy. Let us introduce the following notation:
\begin{equation}
c^{12}=c_3,\quad c^{23}=c_1,\quad c^{13}=c_2.
\label{thecs}
\end{equation}
Notice that we have $c_1^2+c_2^2+c_3^2=1$.

Let us select two colliding fermions with four-momentum $p^\mu_1=(E,\vec{p}_1)$ and $p^\mu_2=(E,\vec{p}_2)$, respectively, were
\begin{equation*}
\vec{p}_1=E(0,\frac{1}{\sqrt{2}},\frac{1}{\sqrt{2}})\quad\text{and}\quad\vec{p}_1=E(0,\frac{1}{\sqrt{2}},-\frac{1}{\sqrt{2}}),
\end{equation*}
in the laboratory reference frame.

Changing to the zero-momentum reference frame by using the Lorentz transformation in (\ref{zeromomentum}), one gets the following values for  the momenta of the ingoing and outgoing fermions:
\begin{equation*}
\begin{array}{l}
{\tilde{p}^\mu_1=\tilde{E}(1,0,0,1),\quad\tilde{p}^\mu_1=\tilde{E}(1,0,0,-1),\quad \tilde{E}=\frac{E}{\sqrt{2}},}\\[8pt]
{\tilde{p}^\mu_3=\tilde{E}(1,\cos\phi \sin\theta,\sin\phi \sin\theta,\cos\theta),\quad\tilde{p}^\mu_4=\tilde{E}(1,-\cos\phi \sin\theta,-\sin\phi \sin\theta,-\cos\theta),
}
\end{array}
\end{equation*}
where $\theta\in[0,\pi]$ and $\phi\in[0,2\pi)$ are the polar and azimuth angles with regard to the direction defined by the three-vector  $(\tilde{p}^1_1,\tilde{p}^2_1,\tilde{p}^3_1)$.

We shall express the  concurrence, $\Delta$ , in (\ref{concurrence}), in terms of $c_1,\,c_2$ and $c_3$ --in (\ref{thecs}), the energy $E$ of the incoming photons in the laboratory reference frame and the angles $\theta$ and $\phi$, which give the direction of the outgoing momenta in the zero-momentum frame --see (\ref{outmomenta}). To do so, the following results will be useful
\begin{equation*}
\begin{array}{l}
{p_1\omega p_2=\cfrac{1}{\Lambda^2_{\text{nc}}}\,p_1^\mu c_{\mu\nu}p_2^\nu=\cfrac{1}{\Lambda^2_{\text{nc}}}\,\tilde{p}_1^\mu\tilde{c}_{\mu\nu}\tilde{p}_2^\nu
=-x\,c_1,}\\[8pt]
{p_1\omega p_3=\cfrac{1}{\Lambda^2_{\text{nc}}}\,p_1^\mu c_{\mu\nu}p_3^\nu=\cfrac{1}{\Lambda^2_{\text{nc}}}\,\tilde{p}_1^\mu\tilde{c}_{\mu\nu}\tilde{p}_3^\nu=}\\
{\phantom{p_1\omega p_3=\cfrac{1}{\Lambda^2_{\text{nc}}}\,p_1^\mu c_{\mu\nu}p_2^\nu=}
-\frac{1}{2}\,x\,[(c_2 + c_3) \sin\phi \sin\theta + c_1 (1 - \cos\theta + \sqrt{2} \sin\phi \sin\theta)],}\\[8pt]
{p_2\omega p_3=\cfrac{1}{\Lambda^2_{\text{nc}}}\,p_2^\mu c_{\mu\nu}p_3^\nu=\cfrac{1}{\Lambda^2_{\text{nc}}}\,\tilde{p}_2^\mu\tilde{c}_{\mu\nu}\tilde{p}_3^\nu=}\\
{\phantom{p_1\omega p_3=\cfrac{1}{\Lambda^2_{\text{nc}}}\,p_1^\mu c_{\mu\nu}p_2^\nu=}
+\frac{1}{2}\,x\,[(c_2 - c_3) \sin\phi \sin\theta + c_1 (1 + \cos\theta + \sqrt{2} \sin\phi \sin\theta)].}
\label{psomegas}
\end{array}
\end{equation*}
where $x$ is the ratio
\begin{equation*}
x=\frac{E}{\Lambda_{{nc}}}.
\end{equation*}
Hence, ${\cal C}_1$ and ${\cal C}_2$ in (\ref{caligraphicces}) are given by the involved expressions
\begin{equation*}
\begin{array}{l}
{{\cal C}_1=-4e^2\sin\big[\frac{1}{4}\,x^2\,[(c_2 + c_3) \sin\phi \sin\theta + c_1 (1 - \cos\theta + \sqrt{2} \sin\phi \sin\theta)]\big]\times}\\
{\phantom{{\cal C}_1=-4e^2}
\sin\big[\frac{1}{4}\,x^2\,[c_1(\cos\theta-1)+(\sqrt{2}\,c_1+c_2-c_3)\sin\theta \sin\phi]\big],}\\[8pt]
{{\cal C}_2=4e^2\sin\big[\frac{1}{4}\,x^2\,[(c_2 - c_3) \sin\phi \sin\theta + c_1 (1 + \cos\theta + \sqrt{2} \sin\phi \sin\theta)]\times}\\
\phantom{{\cal C}_1=4e^2}
{\sin\big[\frac{1}{4}\,x^2\,[c_1(1+\cos\theta)-(\sqrt{2}\,c_1+c_2+c_3)\sin\theta \sin\phi]\big].
}
\end{array}
\end{equation*}

We have seen that in tree level scattering of photons and the head-on  the scattering of fermions in the laboratory reference frame maximal entanglement is achieved if, and only if, $\theta=\pi/2$, and this, whatever the value of the azimuth angle $\phi$. Let see that for the fermion scattering at right angles we are studying the situation  changes drastically. Let us choose $\omega^{\mu\nu}$ so that $c_1=1/\sqrt{3}, c_2=-1/\sqrt{3}$ and $c_3=1/\sqrt{3}$.\footnote{This choice is no loss of generality since we can always rotate appropriately the matrix $\omega^{ij}$ in the action in (\ref{action}).}  Then, if $\theta=\pi/2$, we have
\begin{equation*}
\begin{array}{l}
{{\cal C}_1(x,\phi)=-4e^2\,\sin\big[x^2\,\frac{1 + \sqrt{2} \sin\phi}{4\sqrt{3}}\big]\,\sin\big[x^2\,\frac{-1 +(-2+ \sqrt{2}) \sin\phi}{4\sqrt{3}}\big]}\\[8pt]
{{\cal C}_2(x,\phi)=-4e^2\,\sin\big[x^2\,\frac{-1 + \sqrt{2} \sin\phi}{4\sqrt{3}}\big]\,\sin\big[x^2\,\frac{1 +(-2+ \sqrt{2}) \sin\phi}{4\sqrt{3}}\big].
}
\end{array}
\end{equation*}
From these ${\cal C}_1(x,\phi)$ and ${\cal C}_2(x,\phi)$, by using (\ref{ferconcurrence}),  one gets the following concurrence
\begin{equation*}
\Delta(x,\phi)=\frac{2|{\cal C}_1(x,\phi){\cal C}_2(x,\phi)|}{({\cal C}_1(x,\phi))^2+({\cal C}_2(x,\phi))^2}.
\end{equation*}

It is plain that, whatever the value of $x$, ${\cal C}_1(x,\phi)$ has two zeros corresponding to  the two solutions of
\begin{equation*}
\sin\phi=-\frac{1}{\sqrt{2}}.
 \end{equation*}
These two solutions are $\phi_1=3.92699$ and $\phi_2=5.49779$.
Analogously, the two solutions of
\begin{equation*}
\sin\phi=\frac{1}{\sqrt{2}}
\end{equation*}
gives two zeros of ${\cal C}_1(x,\phi)$, irrespective of the value of x. The solutions of the previous equation read $\phi_3=0.785398$ and $\phi_4=2.35619$.

Hence, we conclude that for any value $x=E/\Lambda_{\text{nc}}$ the concurrence at
$\theta=\pi/2$ has four zeros --i.e., there is no entanglement-- corresponding to the values of $\phi$ given by $\phi_1$, $\phi_2$, $\phi_3$ and $\phi_4$, which are given in the previous paragraph. It can be seen that these are all the zeros of the concurrence if $x< x_c$,
\begin{equation*}
x_c=2\times 6^{1/4}\sqrt{\frac{\pi}{2+\sqrt{2}}}=3.0026...
\end{equation*}
At $x_c$ new zeros of the concurrence begin to pop up.

The reader may find in Figures 1 to 6 plots of the concurrence $\Delta$ --displayed along the vertical axis-- against $\phi$ for the several values of $x=E/\Lambda_{\text{nc}}$. It would appear that, for each value of $\phi$, there is very little variation in the value of $\Delta$ if $x\leq 1$. We believe this is partially due to the fact that a small $\delta x$
variation gives rise to a variation of $\Delta$ which is proportional to $\delta x$ and the zeros of $\Delta(x,\phi)$ do not depend on $x$ if $x\leq 1$. In Figure 7, we plot the difference $\Delta(x=1,\phi)-\Delta(x=10^{-7},\phi)$.
\begin{figure}
\centering
\begin{minipage}{.45\linewidth}
  \centering
  \includegraphics[width=\linewidth]{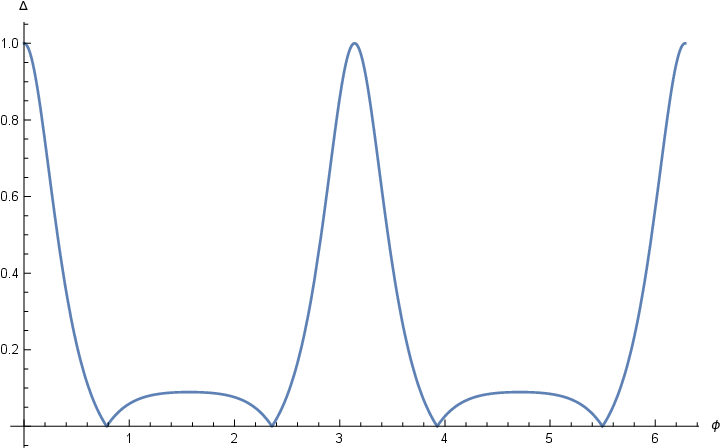}
  \caption{$x=10^{-7}$ and $\theta=\pi/2$.}
\end{minipage}
\quad
\begin{minipage}{.45\linewidth}
  \centering
  \includegraphics[width=\linewidth]{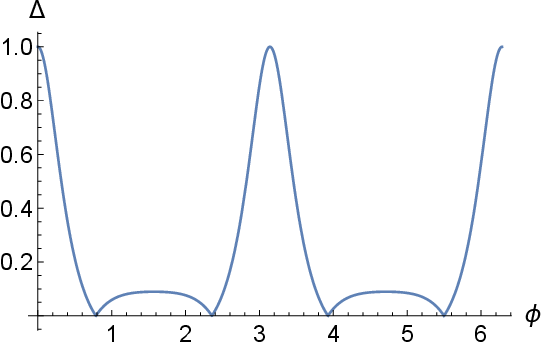}
  \caption{$x=10^{-3}$ and $\theta=\pi/2$.}
\end{minipage}
\begin{minipage}{.50\linewidth}
  \centering
  \includegraphics[width=\linewidth]{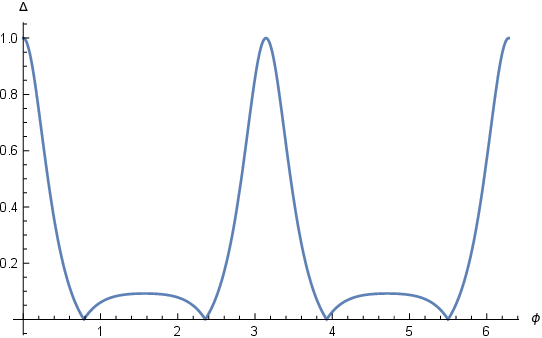}
  \caption{$x=1$ and $\theta=\pi/2$.}
\end{minipage}
\quad
\begin{minipage}{.45\linewidth}
  \centering
\includegraphics[width=\linewidth]{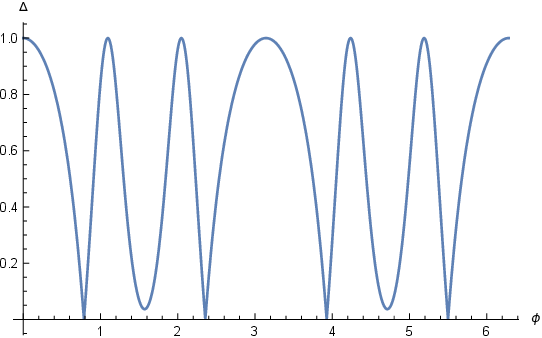}
  \caption{$x=3$ and $\theta=\pi/2$.}
\end{minipage}
\quad
\begin{minipage}{.45\linewidth}
  \centering
  \includegraphics[width=\linewidth]{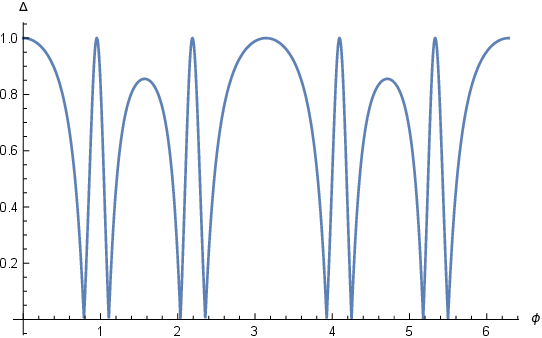}
  \caption{$x=3.1$ and $\theta=\pi/2$.}
\end{minipage}
\begin{minipage}{.45\linewidth}
  \centering
  \includegraphics[width=\linewidth]{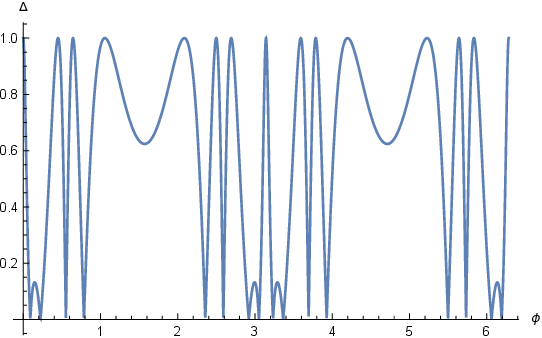}
  \caption{$x=5$ and $\theta=\pi/2$.}
\end{minipage}
\quad
\begin{minipage}{.50\linewidth}
  \centering
  \includegraphics[width=\linewidth]{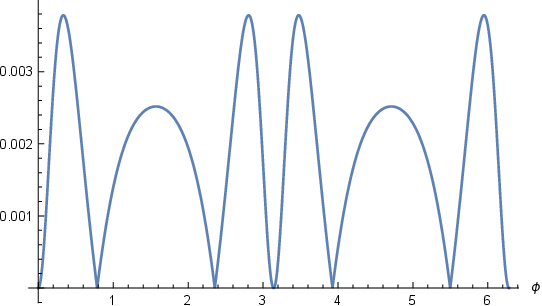}
  \caption{$\Delta(1,\phi)-\Delta(10^{-7},\phi)$.}
\end{minipage}
\end{figure}

\newpage

\section{Conclusions}

Entanglement may also be generated through tree-level scattering in noncommutative gauge theories even in the event that the scattering process in question did not exist in ordinary Minkowski spacetime. In some cases the amount of scattering generated does not depend on the value of the noncommutative matrix $\omega^{\mu\nu}$ and in other instances it does. We have seen that in noncommutative quantum electrodynamics the scattering of photons with opposite helicity has, at tree-level, a concurrence which is the same as in the scattering of gluons in ordinary Minkowski spacetime. The same result is obtained when zero-charge fermions collide  head-on in the laboratory reference frame; this is a  tree-level scattering process which does not exist in ordinary Minkowski, as does not exist the scattering of photons. Of course, in both these cases maximal entanglement is obtained if and only if $\theta$, the polar angle in the zero-momentum frame, is $\pi/2$.
However, we have exhibited a scattering process where the concurrence depends on the energy of the colliding fermions, the noncommutativity matrix, $\omega^{\mu\nu}$, and the polar, $\theta$, and azimuth, $\phi$, angles of the zero-momentum frame. In this scattering process the fermions have opposite helicity and collide at right angles in the laboratory reference frame. Here, when $\theta=\pi/2$, there are always values of $\phi$ for which there is no entanglement.

\section{Acknowledgements.}
This piece of research work has been  financially supported  in part by the Spanish Ministry of Science, Innovation and Universities through grant PID2023-149834NB-I00.

\end{document}